\documentclass[]{jd04}    
\usepackage[]{graphicx}

\Title[OB Spectra]{Multiwavelength Systematics of OB Spectra} 
\Author{Walborn$^1$, Nolan R.} 
\Address{$^1$Space Telescope Science Institute,\\
3700 San Martin Drive, Baltimore, Maryland 21218, USA\\}{}
\ShortAuthors{Walborn}


\begin{document}
\maketitle

\begin{abstract}
The systematics of OB spectra are reviewed in the optical domain,
dominated by photospheric lines, and in the far ultraviolet (both {\it IUE} 
and {\it FUSE} ranges), in which the stellar-wind profiles dominate.  First, 
the two-dimensional (temperature, luminosity) trends in normal spectra 
are surveyed.  Then, the normal reference frame having been established,
various categories of peculiar objects can be distinguished relative to
it, which reveal several phenomena of structural and/or evolutionary
significance.  Included are CNO anomalies at both early and late O 
types, three varieties of rapid rotators, hot and cool Of/WN transition 
objects, and the recently discovered second known magnetic O star.  
The importance of both optical and UV observations to understand these 
phenomena is emphasized; for instance, progress in understanding the 
structure of the new O-type magnetic oblique rotator is hampered by the 
current lack of a UV spectrograph.  While progress in the physical 
interpretation of these trends and anomalies has been and is
being made, increased attention to modeling the systematics would
accelerate future progress in this author's opinion.  Finally,
preliminary results from a {\it Chandra} high-resolution survey of OB X-ray
spectra (PI W. Waldron) are presented.  They provide evidence that, just
as emerged earlier in the UV, systematic morphological trends exist in
the X-ray domain that are correlated with the optical spectral types, and
hence the fundamental stellar parameters, contrary to prevailing opinion.
\end{abstract}

\section{Introduction}

The Morgan-Keenan (MK) System of spectral classification, one of the 
foundations of stellar astrophysics, is a classical application of
morphological techniques.  A reference frame of standard spectra, with
empirical line-ratio criteria therein, is established.  Then new spectra
are described differentially relative to the standards, with
observational parameters as similar as possible (preferably identical) to
those of the standards.  Normal spectra are classified into the system,
while peculiar exceptions to the standard behavior of the criteria  
may also be recognized.  Both in the definition of the system and 
especially in its application, independence from external information is
paramount, including even physical calibration and interpretation of the
data themselves.  In this way, errors and uncertainties in the subsequent
procedures do not affect the description of the phenomena, which remains
valid in the event of revisions or improvements to the latter; and
correlations or discrepancies with other kinds of data may be usefully
investigated.  The hazards of ignoring these precepts, whenever a diverse
phenomenology is beyond the capacity of current models to accurately
predict and explain, can be readily appreciated with the benefit of
hindsight in a collection of specialist essays on a new system of spectral 
classification edited by Schlesinger (1911).    

A current view holds that astrophysics has rendered astronomical
morphology obsolete.  However, when a new observational domain is opened
to investigation, such as a new wavelength regime, a different
metallicity in an external galaxy (or region of our own), or simply
increased information content in higher quality data, the above
principles apply fully.  Stated another way, an adequate image of the new
phenomena must be formulated before they can usefully be subjected to
interpretation or modeling.  Analogous examples of confusion can be found
in the more recent literature, e.g., on the relationship of OB stellar
winds to the fundamental stellar parameters during the early 1980's.
This paper presents specific, practical examples of systematic trends and
relationships, as well as peculiar phenomena, recently discovered in OB
spectra by the application of morphological techniques.  It is also
suggested that systematic modeling relative to an analogous reference frame 
of standard objects could accelerate progress toward the ultimate objective
of physical understanding. 

\section{Reference Frame}

\subsection{Spectral Type}

The optical O-type horizontal (temperature) classification is based upon 
the helium ionization, primarily the absorption-line ratio of 
He~II~$\lambda$4541/He~I~$\lambda$4471, which has a value of unity at
spectral type O7 (e.g., Walborn \& Fitzpatrick 1990; WF).  The He~I line 
is very weak at type O4; type O3 was introduced for several stars in the 
Carina Nebula by Walborn (1971a), based upon the absence of the He~I line in 
the photographic data of the time.  It can often be seen in O3 spectra with 
modern, high-S/N digital data, but the interpretation of the very weak 
feature is compromised by later type companions, imperfect nebular 
emission-line subtraction, and possibly supergiant winds, as well as the 
noise level.  Hence, Walborn et al. (2002a) refined the classification 
at the earliest types on the basis of the 
N~IV~$\lambda$4058/N~III~$\lambda\lambda$4634-40-42 {\it selective
emission-line} ratio (Walborn 2001), adding new spectral types O2 and O3.5
to accommodate its observed range.  A sequence of very early-O supergiant
spectra is shown in Figure~1.

\begin{figure}
\centerline{\includegraphics[width=110mm]{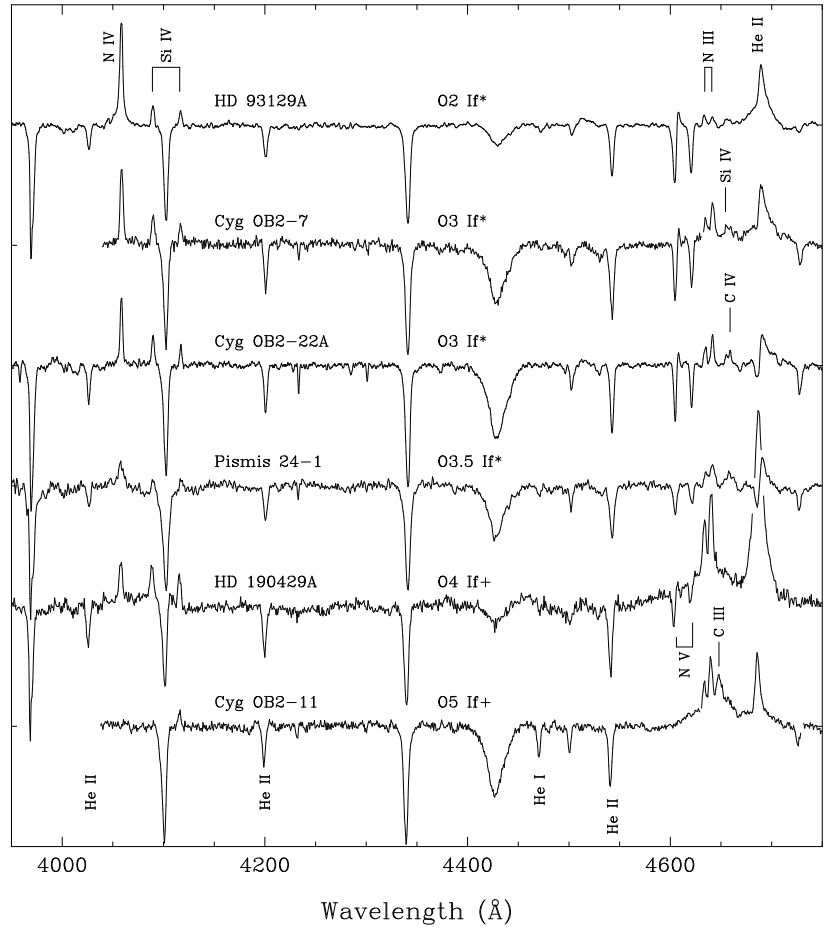}}
\caption{Temperature sequence of very early O blue-violet spectra.
Courtesy of Ian Howarth.}
\label{fig1}
\end{figure}

The original, photographic definition of the transition from type O9.5 to
B0 was the disappearance of He~II~$\lambda$4541, although it is clearly
seen at the latter type in modern data.  On the B main sequence (luminosity 
class V), the behavior of the entire blue-violet He~I spectrum, which has 
maximum intensity at type B2, is the primary spectral-type indicator.  
In the supergiants, the silicon ionization, in particular
Si~IV~$\lambda$4089/Si~III~$\lambda$4552 at the earlier types and the
Si~III relative to Si~II$\lambda\lambda$4128-30 at mid-B types, provides
the primary horizontal criterion.  At late-B types, the strengthening of
Mg~II~$\lambda$4481 relative to the declining He~I~$\lambda$4471 is a
useful criterion.  CNO lines provide additional supporting criteria in
(morphologically) normal spectra, but must be used with caution in view
of the anomalies they can display in some spectra (Section 3.1 below).  
Digital sequences can be found in WF.

Large-scale atlases of high-resolution {\it International Ultraviolet
Explorer} ({\it IUE}) data show the very tight correlations of the UV
stellar-wind profiles with both O and B spectral types in the great
majority of normal spectra (Walborn et al. 1985, 1995, respectively).
On the main sequence, the N~V~$\lambda\lambda$1239-43 and
C~IV~$\lambda\lambda$1548-51 resonance lines have broad, saturated
P~Cygni profiles through type O6 and decline smoothly thereafter, while
Si~IV~$\lambda\lambda$1394-1403 shows no wind effect anywhere on the main
sequence.  In the supergiants, on the other hand, there is an
O~V~$\lambda$1371 wind profile at types O2-O3, while Si~IV has a weak
wind profile at O4, which grows to a maximum at mid/late-O and declines
thereafter through the B sequence.  In the latter,
C~II~$\lambda\lambda$1334-36 and Al~III~$\lambda\lambda$1855-63 develop
wind profiles with maxima at types B1-B2.  The {\it Far Ultraviolet 
Spectroscopic Explorer} ({\it FUSE}) allowed this systematic phenomenology 
to be extended to numerous additional species and ionizations in the
900-1200~\AA\ range, including the superionized 
O~VI~$\lambda\lambda$1032-1038 (Walborn et al. 2002b, Pellerin et al. 2002).

\subsection{Luminosity Class}

The MK System contained no vertical (luminosity) classification for
spectra earlier than type O9.  Such a system was introduced by Walborn
(1971b, 1972, 1973), based upon identification of the Of phenomenon
(selective emission effects in He~II~$\lambda$4686 and 
N~III~$\lambda\lambda$4634-40-42) with negative luminosity effects in
absorption (i.e., decreasing absorption strength with increasing
luminosity, inferred to be caused by emission filling) in the same lines
at types O9-B0.  A luminosity sequence of blue-violet optical spectra at
spectral type O6.5 is shown in Figure~2.  It can be seen that
He~II~$\lambda$4686 is a strong absorption on the main sequence (class V), 
which weakens gradually through the giants and comes into emission in the
Ia supergiant, while the accompanying N~III emission strengthens
correlatively.  These sequential configurations are denoted by ((f)),
(f), and f in the spectral types, as labeled in the figure.  Calibrations
in terms of absolute visual magnitude corroborated these morphological
classifications, and the N~III temperature and gravity dependence was
reproduced theoretically by Mihalas et al. (1972).

\begin{figure}
\centerline{\includegraphics[width=150mm]{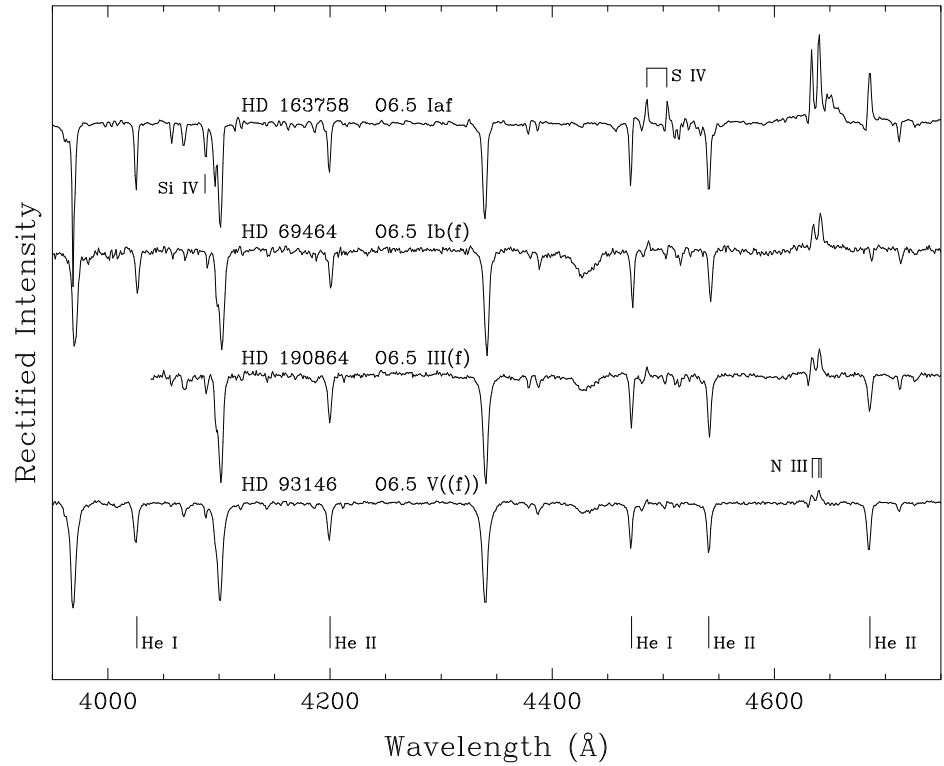}}
\caption{Luminosity sequence of mid-O blue-violet spectra.  Courtesy of
Ian Howarth.}
\label{fig2}
\end{figure}

An ``inverse Of effect'', i.e. He~II~$\lambda$4686 absorption {\it stronger}
relative to the other He lines than in class V spectra, usually found in
very young regions, has been hypothesized to correspond to lower (visual)
luminosities and smaller ages.  That is, typical class V spectra may
already have some emission filling in that line, which is less or absent in
these ``Vz'' spectra.  Some examples in the Large Magellanic Cloud H~II
region N11 are shown in Figure~3 (see Walborn \& Parker 1992, Parker et
al. 1992).  These objects may be near or on the zero-age main sequence (ZAMS),
contrary to some expectations that such would not be optically observable at
high masses.  This topic has been reviewed by Walborn (2006), in which
fifty morphologically selected candidate ZAMS O stars are listed.  It is
a promising subject for future astrophysical investigation.

\begin{figure}[t]
\hskip -4mm
\centerline{\includegraphics[angle=90,width=140mm]{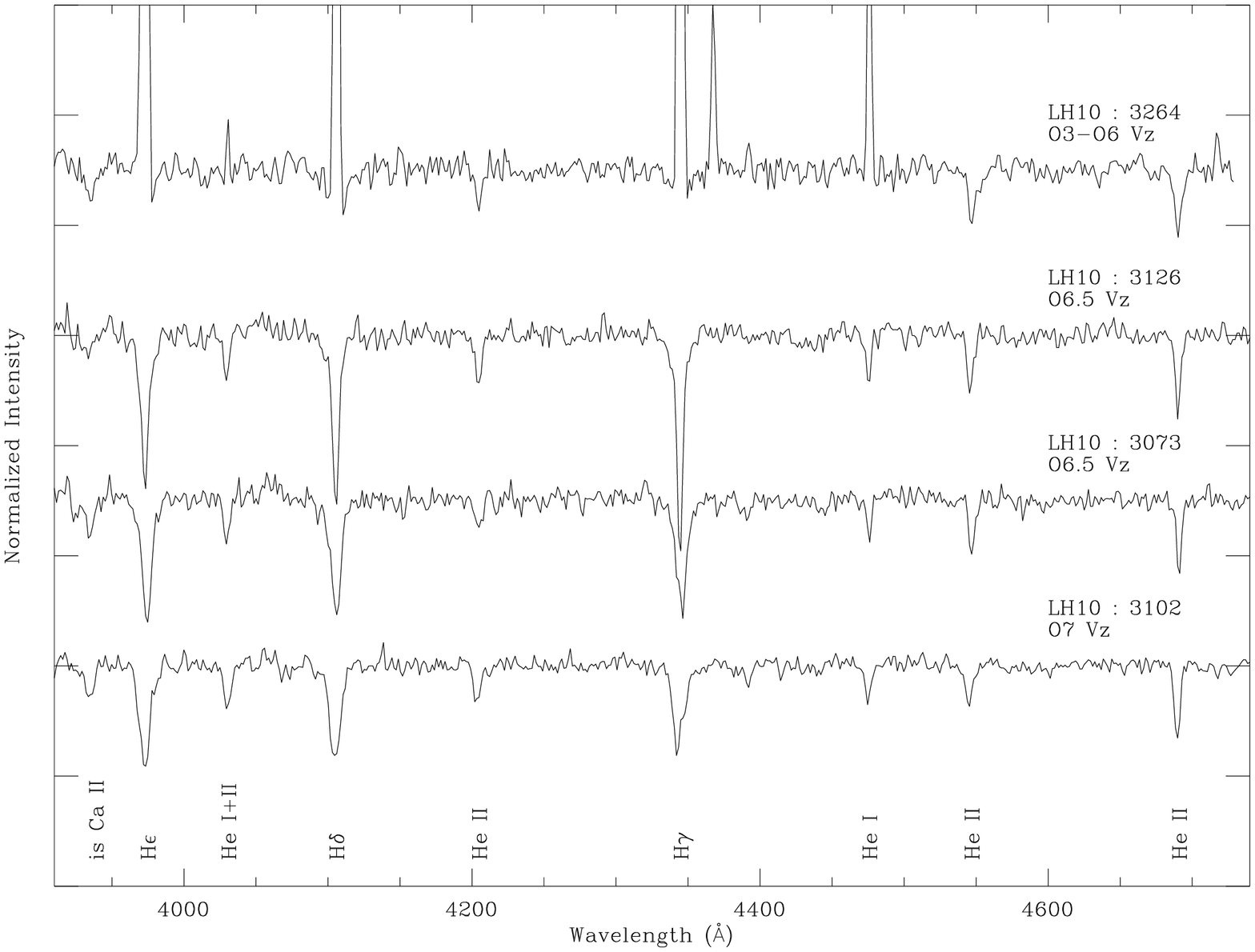}}
\caption{Vz blue-violet spectra in Lucke-Hodge 10, Henize N11 in the LMC.
Courtesy of Joel Parker.}
\label{fig3}
\end{figure}

The UV wind spectra display remarkable correlations with the optical
luminosity classes.  In particular, in the {\it IUE} data, the Si~IV
resonance line progresses smoothly from no wind effect on the main
sequence, through intermediate wind profiles in the giants, to a fully
developed P~Cyg profile in the Ia supergiants; Figure~4 shows the UV
spectra of the same stars displayed optically in Figure~2 (Walborn \&
Panek 1984a, Walborn et al. 1985).  This is essentially an ionization
effect: the Si~IV potential of 45~eV is significantly less than those of
N~V (98~eV) and C~IV (64~eV), which allows the former to respond to the
density range among these winds, while the latter two remain saturated
throughout.  The {\it FUSE} range offers three more similarly
luminosity-sensitive features: C~III~$\lambda$1176, 48~eV;
S~IV~$\lambda\lambda$1063-73, 47~eV; and P~V~$\lambda\lambda$1118-28,
65~eV.  The last of these retains density/luminosity sensitivity because
of the {\it very low abundance} of P, rather than the ionization
potential (Walborn et al. 2002b). 

\begin{figure}[t]
\hskip 7mm
\centerline{\includegraphics[angle=90,width=140mm]{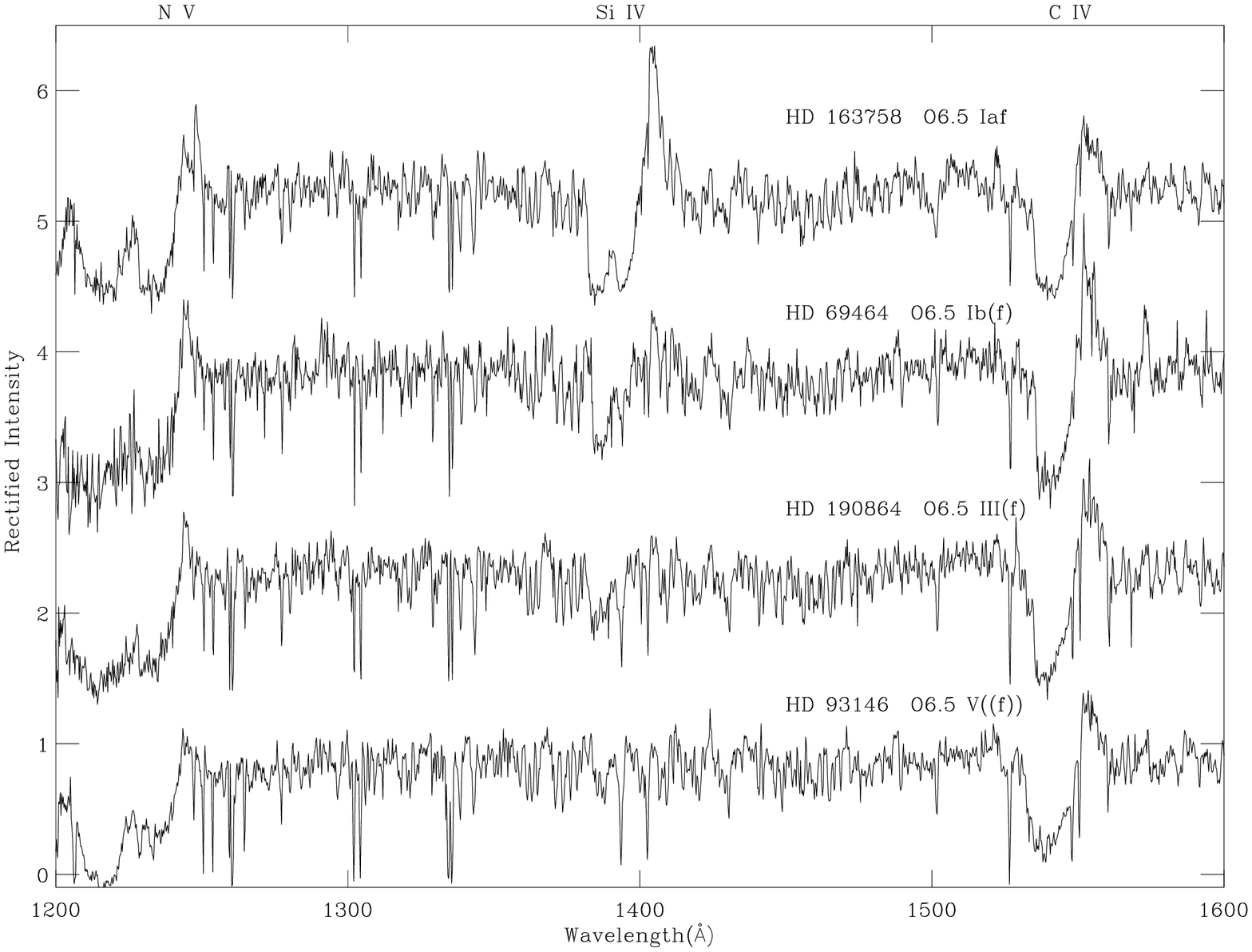}}
\caption{Luminosity sequence of mid-O FUV ({\it IUE}) spectra (the same
stars as shown in Fig.~2).  Courtesy of Danny Lennon.}
\label{fig4}
\end{figure}

The yellow-red region of OB spectra contains several interesting
diagnostic lines (Walborn 1980).  In preparing this review, the author
found a very sensitive temperature/luminosity effect in the selective
emission line C~III~$\lambda$5696: on the main sequence at type O9 it is
absent (neutralized), but at O9.5~V it is a weak absorption line, which
then weakens further and comes into emission with increasing luminosity, 
as shown in Figure~5.  When all of these detailed effects are reproduced 
by the models, the definition of the physical parameters will be highly
constrained.     

\begin{figure}
\centerline{\includegraphics[width=140mm]{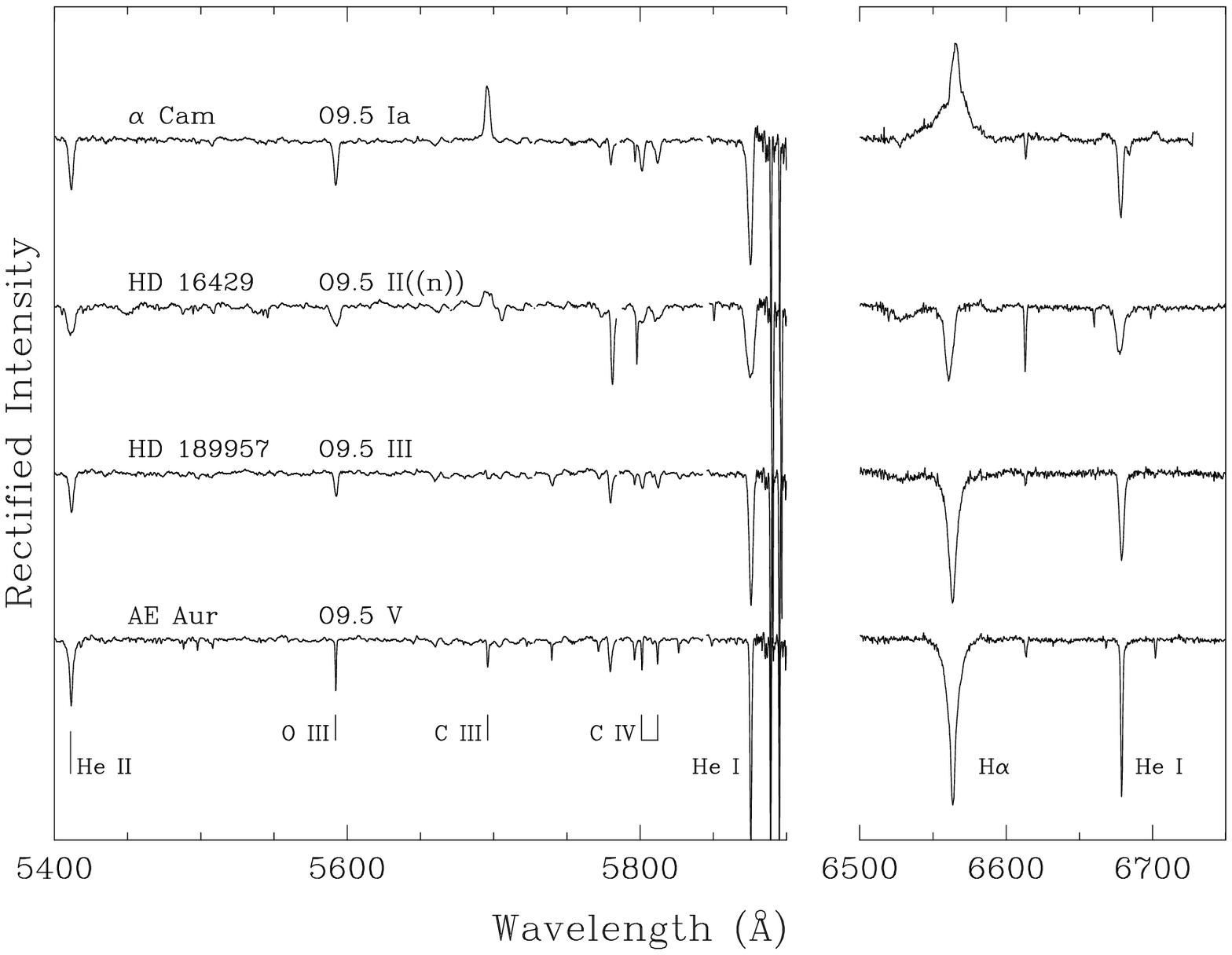}}
\caption{Luminosity sequence of late O yellow-red spectra.  Courtesy of
Ian Howarth.}
\label{fig5}
\end{figure}

Luminosity classes in the B-type range depend primarily on Si/He line
ratios, Stark effects in certain He~I lines, and secondarily on the behavior
of CNO lines, again used with caution.  See WF for illustrative sequences
in the optical, and Walborn et al. (1995) for correlative effects in the UV.

\section{Peculiar Categories}

\subsection{CNO Anomalies}

A review of inverse CNO anomalies in OB absorption-line spectra, denoted
as OBN and OBC, was given by Walborn (1976), and an update by Walborn
(2003).  It is now generally accepted that the morphologically normal
majority of OB supergiants display an admixture of CNO-cycled material in
their atmospheres and winds, while the relatively rare OBC objects have
physically normal (i.e., main-sequence) CNO abundances, and the OBN have
more extreme mixing as a result of either binary interactions or rapid
initial rotational velocities, with homogeneous evolution in extreme cases
(Maeder \& Meynet 2000).  The optical anomalies are usually reflected in
the UV wind profiles (Walborn et al. 1985, 1995).

The recent discovery of a CNO dichotomy among O2 giants in the Magellanic
Clouds, initially from a survey of the 3400~\AA\ region in their spectra
(Walborn et al. 2004a), was a surprise.  These very massive objects 
have small absolute ages and lie near the main sequence, indicating more
rapid mixing processes than contemplated in current models, and/or very
rapid initial rotations perhaps inducing homogeneous evolution back toward 
the main sequence.  They represent a challenge to the models and ultimately 
a powerful diagnostic of early massive stellar evolution.  Further related 
results from the 3400~\AA\ survey are presented by Morrell et al. (2005).

Typical CNO anomalies in the optical spectra of late-O supergiants are
illustrated in Figure~6.  More detailed descriptions of these
high-quality data may be found in Walborn \& Howarth (2000), including
identifications of the plethora of weak CNO lines that faithfully track 
the inverse ON/OC dichotomy.

\begin{figure}
\centerline{\includegraphics[width=150mm]{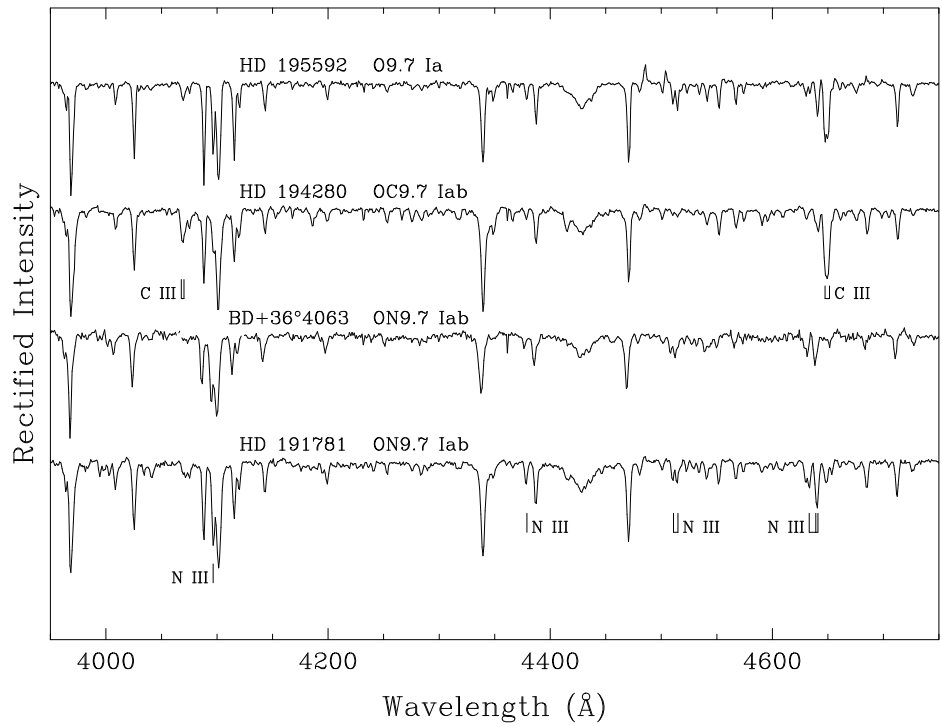}}
\caption{CNO anomalies in late O supergiant blue-violet spectra.  Courtesy of
Ian Howarth.}
\label{fig6}
\end{figure}
 
\subsection{Rapid Rotators}

Three varieties of O-type rapid rotators are illustrated in Figure~7.
HD~155806 is a relatively rare analogue of the Be stars; its yellow-red
spectrum including H$\alpha$ is reproduced by Walborn (1980).  See also
Negueruela et al. (2004) for a comprehensive discussion of the Oe class.  
HD~191423 is one of the most rapidly rotating stars known and a prototype
of the ONn class, which is directly relevant to enhanced mixing of
processed material in rapid rotators (Howarth \& Smith 2001; Walborn
2003; Howarth 2004).

\begin{figure}[t]
\centerline{\includegraphics[angle=270,width=140mm]{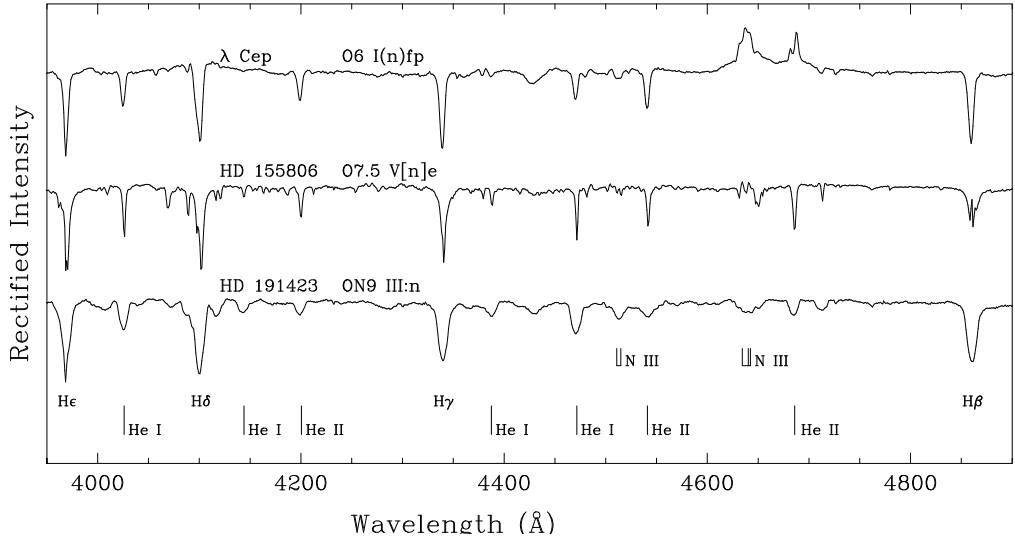}}
\caption{Blue-violet spectra of three kinds of O-type rapid rotators.  
Courtesy of Ian Howarth.}
\label{fig7}
\end{figure}

Another intriguing class is the Onfp (Walborn 1973; or Oef, Conti \& Leep
1974), represented by $\lambda$~Cephei in Figure~7.  These spectra have
comparable broadening in absorption lines and Of emission features, with
a prominent absorption reversal in the He~II~$\lambda$4686 emission line
that may indicate the presence of a hot disk.  Numerous luminous members 
of this class are being found in the Magellanic Clouds (Walborn et al.
2000, A.~Moffat et al. in preparation, P.~Crowther et al. in preparation,
I.~Howarth et al. in preparation).  These objects are interesting
candidates for stellar merger remnants and/or gamma-ray burst progenitors. 
  
\subsection{``Slash'' Stars}

Two categories of O-type spectra with prominent emission lines, which are
related to the WN sequence, were given composite or dual classifications 
that have led to their being referred to as ``slash'' stars.  The hotter 
category is evidently intermediate between very hot Of and luminous WNL 
spectra and is found associated with those two classes in giant H~II 
regions such as 30~Doradus.  These objects retain prominent very early 
O-type absorption spectra, but they have stronger winds than pure Of 
stars, that produce emission features of intensity and width more
similar to those in WN spectra.  High-quality optical and UV observations 
of a prototypical example, Melnick~42 in 30~Dor, are compared with related 
Of and WN spectra by Walborn et al. (1992).  This category likely represents 
the transition between Of and WNL phases of the most massive stars.

A cooler category of ``intermediate'' spectra was isolated in the LMC and
designated Ofpe/WN9 (Walborn 1977, 1982; Bohannan \& Walborn 1989).
These objects, together with Galactic extreme O~Iafpe stars, were 
subsequently reclassified into a WN9-11 (WNVL) sequence by Crowther \& Smith 
(1997); see also Crowther \& Bohannan (1997) and Walborn \& Fitzpatrick 
(2000).  In the UV, most of these objects display relatively low-ionization,
shortward-shifted absorption features, indicative of very dense,
low-velocity winds (Pasquali et al. 1997).  One of the original
prototypes of this category, HDE~269858 or Radcliffe 127, entered a
classical Luminous Blue Variable outburst state in 1982 (Stahl et al. 1983).
Other category members have LBV-like, axisymmetric, N-rich circumstellar
nebulae (Walborn 1982, Nota et al. 1995, Pasquali et al. 1999), evidently
ejected in prior events.  Thus, some or all of these objects correspond
to quiescent phases of LBVs, an important insight into the still
mysterious, rapid transitions during the late evolution of massive stars.
It is plausible that they may subsequently reach WNE and/or WC states,
following the extensive LBV mass loss.

\subsection{Magnetic Stars}

The Of?p designation was introduced by Walborn (1972) to distinguish 
the peculiar spectra of HD~108 and HD~148937 from normal Of spectra.
The question mark was intended to emphasize that these objects were not
believed to be normal Of supergiants, as the latter had just been
interpreted.  Walborn (1973) added a third member to this class, 
HD~191612.  The defining peculiarity in the blue spectra is C~III 
$\lambda\lambda$4647-4650-4651 emission lines of comparable intensity to 
N~III $\lambda\lambda$4634-4640-4642; the former are usually much weaker 
than the latter when present at all in normal Of spectra.  Other 
line-profile peculiarities in the Of?p spectra are suggestive of shell 
phenomena or dilute (circumstellar) material.  Subsequent {\it IUE}
observations confirmed that these stars are not supergiants, in terms of
the behavior of the Si~IV resonance lines.  Spectral variations had 
been reported in HD~108 and have been well documented by Naz\'e et al. 
2001; the C~III/N~III emission-line ratio and emission components 
at H and He lines change on a timescale of decades.  The spectral 
(in)stability of HD~148937 is less known, but it is surrounded by 
spectacular axisymmetric, N-rich ejected nebulosities (NGC~6164-6165), 
reminiscent of Luminous Blue Variable nebulae (references in Walborn 
et al. 2003).

Interest in HD~191612 was rekindled in 2001 when it was realized that the
spectrum observed by Herrero et al. (1992) was completely different from
the one published by Walborn (1973).  In particular, the C~III emission
was absent (!), the spectral type was O8 as opposed to O6.5 in the
earlier observation, and the profiles of other features such as He~II 
$\lambda$4686 had changed entirely.  Subsequent literature and archival
searches, together with new observations, documented the recurrence and
strict reproducibility of these spectral variations, although the record
did not allow definitive identification of the timescale (Walborn et al.
2003).  A surprising result was the variation of H$\alpha$ from a strong
P~Cygni profile in the O6 state to predominantly absorption in the O8,
suggesting large changes in the mass-loss rate.  Further observations 
during 2003 and 2004 showed that the recurrent spectral states last less 
than a year, and the {\it Hipparcos} photometry provided the breakthrough 
datum of a 538~d period in a very low-amplitude lightcurve (Naz\'e 2004), 
which satisfies all available spectroscopy since at least 1982 (Walborn et al. 
2004b).  Figure~8 illustrates the drastic phase dependence of the spectrum
in both the blue-violet and yellow-red.  Subsequent analysis of very 
extensive optical data with complete phase coverage demonstrates that
the spectral-type variation is caused by filling in of the He~I lines in 
the O6 state rather than an effective-temperature change, and that the O8 
spectrum, while still peculiar, is the baseline (I. Howarth et al., 
in preparation). 

\begin{figure}
\vskip -3cm
\centerline{\includegraphics[width=140mm]{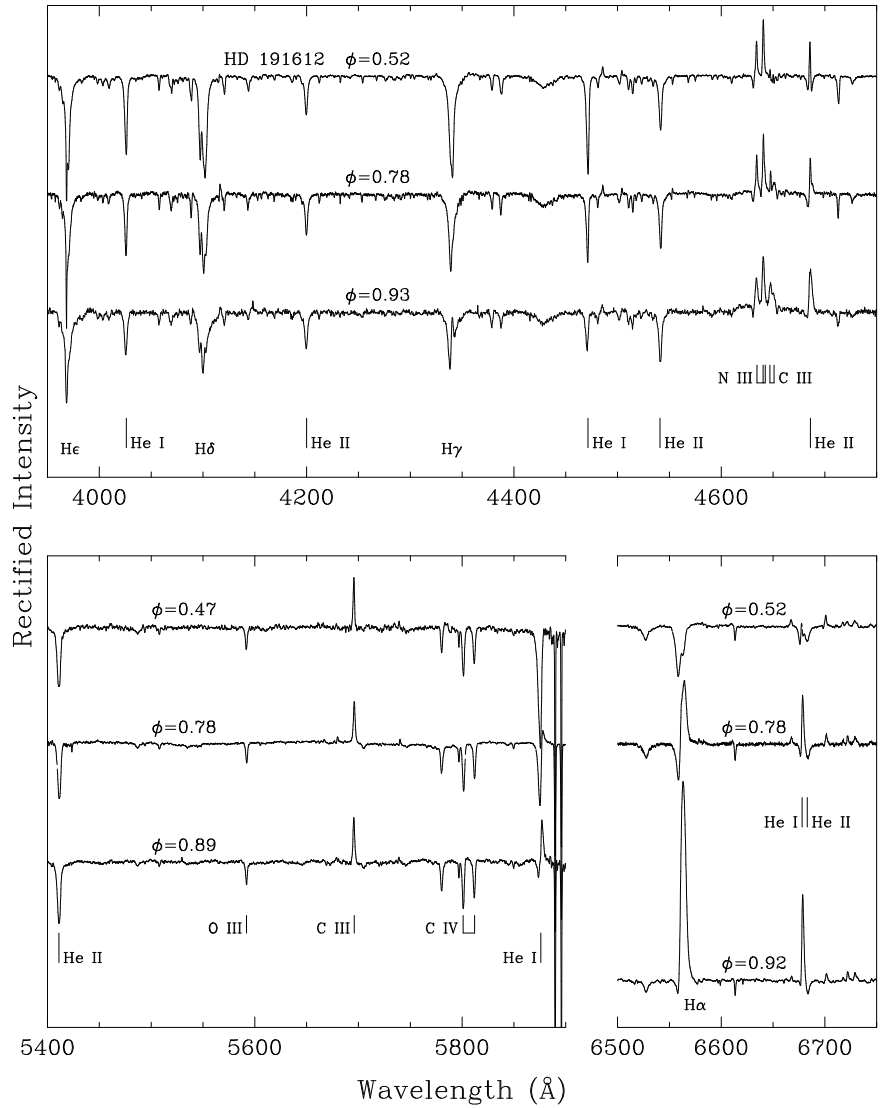}}
\caption{Blue-violet (top) and yellow-red (bottom) spectra of the Of?p
magnetic oblique rotator HD~191612 at different phases of the 538~d
rotational period.  Courtesy of Ian Howarth.}
\label{fig8}
\end{figure}
 
The bizarre phenomena exhibited by HD~191612 are unprecedented in an
O-type star and challenged physical interpretation.  A second breakthrough
is only the second detection of a magnetic field in an O-type star, by
Donati et al. (2006a).  While phase coverage remains to be obtained, this
observation suggests that the variations may be caused by an oblique
rotator configuration with a magnetically confined wind disk, and that
the very long rotational period is a result of magnetic braking.
Comparison with the first known O-type magnetic oblique rotator,
$\theta^1$~Orionis~C, is thus also suggested (Donati et al. 2002, Smith \&
Fullerton 2005, Gagn\'e et al. 2005, Wade et al. 2006).  Although of
similar mass, this star is much younger, consistent with its shorter
period of 15~d.  $\theta^1$~Ori~C displays large, phase-dependent
variations in its UV wind features (Walborn \& Nichols 1994, Stahl et al.
1996), which have provided key diagnostics for the physical models.
There are ongoing attempts to obtain UV spectroscopic phase coverage of
HD~191612 with {\it FUSE}, but current pointing limitations render that
difficult, and the restoration of appropriate capabilities to {\it HST}
depends upon a new successful servicing mission, now planned for 2008.

It is remarkable that all four of the hottest magnetic stars known to
date were isolated as peculiar from their optical and/or UV spectra in
advance of the magnetic detections.  The other two are $\tau$~Scorpii
(Walborn \& Panek 1984b; Walborn et al. 1985, 1995; Donati et al. 2006b)
and $\xi^1$~Canis~Majoris (Rountree \& Sonneborn 1991, 1993; Walborn et al. 
1995; Hubrig et al. 2006).  This circumstance suggests the strong magnetic
candidacy of other OB stars with unexplained spectral peculiarities
and/or variations: in addition to the other two Galactic Of?p stars
above, they are HD~36879 (Walborn \& Panek 1984b, Walborn et al. 1985),
$\theta$~Carinae (Walborn et al. 1995, Lloyd et al. 1995), and
15~S~Monocerotis (unpublished).  

\section{X-Ray Systematics}

The spectroscopic capabilities of the {\it Chandra} (and {\it XMM-Newton})  
X-ray observatories permit for the first time the extension of
morphological techniques as described above in the optical and UV
domains, to the X-ray line spectra of the OB stars.  A {\it Chandra}
program (PI W.~Waldron) to fill gaps in the archival HR Diagram coverage
has been conducted.  Although such coverage to date remains sparse, it is
now sufficient to support a preliminary investigation of the X-ray
spectral systematics in relation to the optical spectral types of the
stars.  To that end, supergiant/(giant) and main-sequence/(giant) X-ray
spectral sequences from {\it Chandra} HETGS data are displayed in
Figures~9 and 10, respectively.  It should be emphasized that these stars
have been selected as normal representatives of their spectral types;
e.g., the magnetic stars discussed in the previous section also have 
peculiar X-ray spectra and must be omitted from the search for
fundamental morphological trends.

\begin{figure}
\vskip -0.5cm
\centerline{\includegraphics[angle=180,width=140mm]{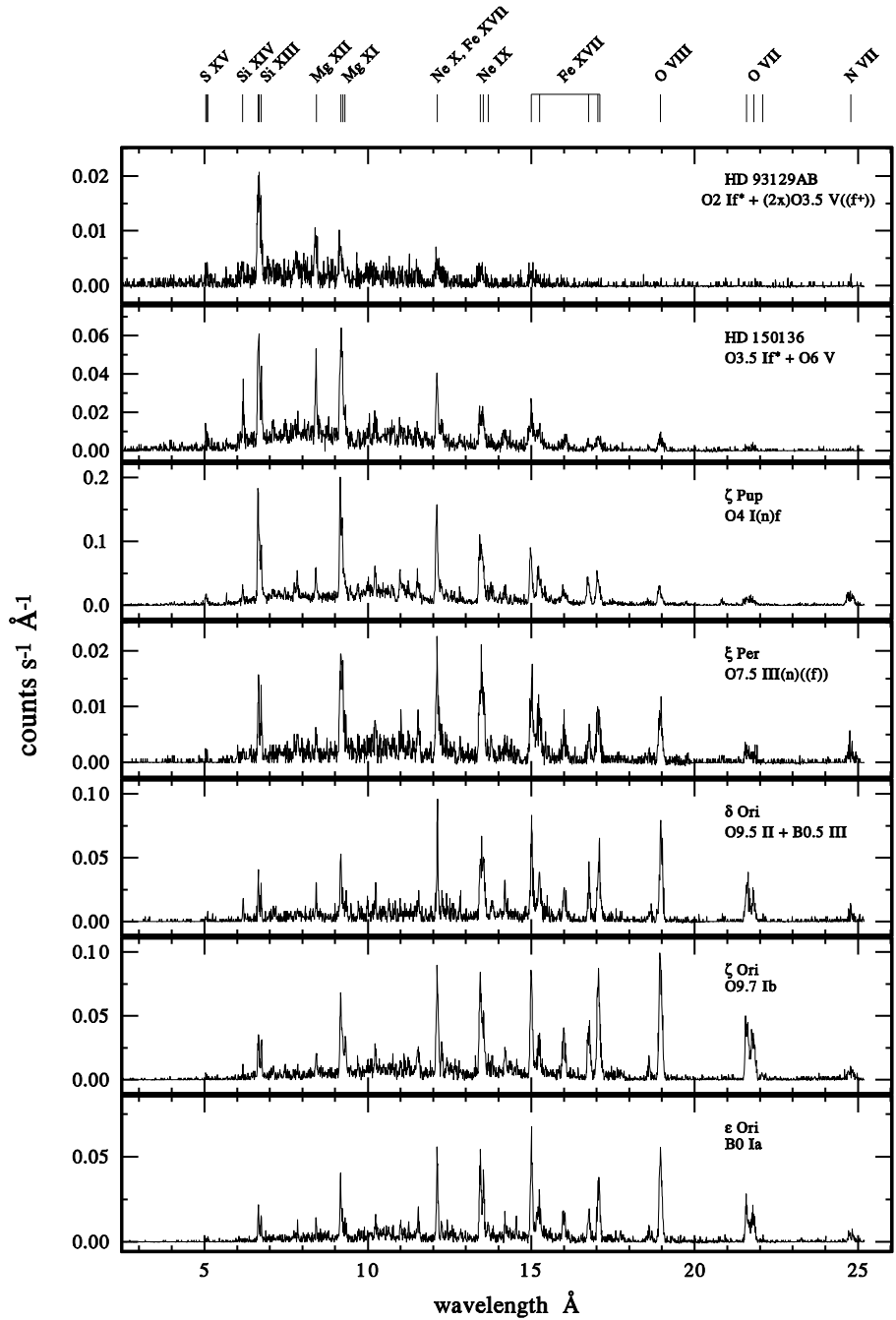}}
\caption{Sequence of OB supergiant/(giant) X-ray spectra from {\it
Chandra}.  Courtesy of Wayne Waldron.}
\label{fig9}
\end{figure}

\begin{figure}
\vskip -0.5cm
\centerline{\includegraphics[angle=180,width=140mm]{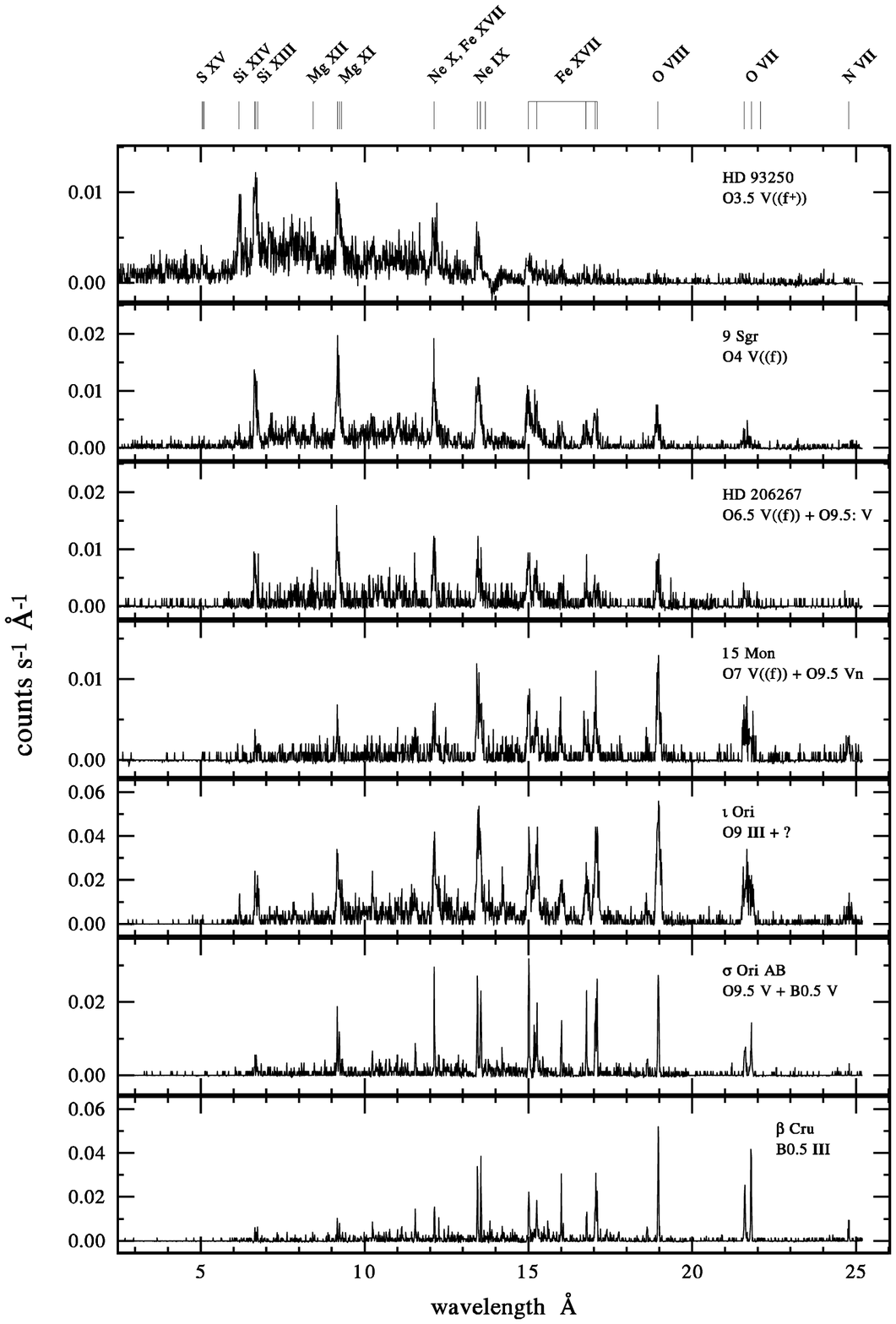}}
\caption{OB main-sequence/(giant) X-ray spectra from {\it Chandra}.  
Courtesy of Wayne Waldron.}
\label{fig10}
\end{figure}

The existence of such trends is readily apparent in the figures.  First,
the strongest lines migrate toward longer wavelengths with advancing
spectral type, which is an ionization effect.  Second, the ratios of the 
close pairs of He- and H-like ionic lines from Si, Mg, Ne, and O display
correlations with the spectral types.  For instance, the rapid declines in
Mg~XII/Mg~XI in the early O supergiants, and of Si~XIV/Si~XIII on the
early O main sequence, are noteworthy.  (The weakness of the Mg~XII line in
the main-sequence spectra may be a luminosity effect, although current
coverage is inadequate to establish that; the weakness of the Si~XIV line 
in HD~93129 is a surprising anomaly for further investigation.)  The
reversal of the Ne~X/Ne~IX ratio in both sequences, despite interference
from Fe~XVII at the later types, is remarkable, as is the smooth decline 
in the absolute strength of the Si~XIII line.  Several of these objects
are believed to be colliding-wind binaries, which nevertheless does not
appear to obstruct the observed trends; neither does the range of extinctions 
among these stars, to which the ratios of close line pairs should be 
particularly insensitive.  We are currently also investigating the 
behavior of detailed line properties such as width, shape, shift, and He-like 
forbidden/intercombination/recombination component ratios along these sequences.

These trends in the X-ray spectra of the OB stars as a function of the 
optical spectral types (and by implication, of the fundamental stellar 
parameters) are unexpected in some views of their origin, and they have 
not emerged from previous studies because of inadequate samples and current 
modeling uncertainties.  In effect, the history of the discovery 
of the UV wind-profile systematics (Walborn et al. 1985, 1995) 
appears to be repeating in the X-ray domain.  The importance of pure
morphological investigation of such trends, as emphasized in the
Introduction, is being demonstrated once again.  Most likely, the
physical origin of these correlations will be found in the winds
themselves; in retrospect, that may not be so surprising in view of known
relationships between bolometric and X-ray luminosities, as recently
demonstrated in detail in NGC~6231 by Sana et al. (2006).  These
morphological results will provide strong guidance to further
developments in physical models of the phenomena.  Progress will likely be
accelerated if astrophysics emulates some of the morphological techniques, 
e.g., by defining standard objects that are homogeneously reanalyzed whenever
there are substantial revisions to the models, and by emphasizing the
modeling of the powerfully diagnostic, relative trends in the HRD, as opposed 
to exclusive, absolute studies of one or a few objects in isolation.

\noindent
{\it Acknowledgements:}
The full array of multiwavelength spectrograms presented in this review 
will be reproduced and described in more detail in the OB chapter of a 
new book on spectral classification being prepared by R.O.~Gray and
C.J.~Corbally.  The author sincerely thanks I.D.~Howarth, D.J.~Lennon, 
P.A.~Crowther, and W.L.~Waldron for providing extensive data and 
spectral plots for this review.  Other, previous plots were made by
A.W.~Fullerton, J.Wm.~Parker, N.I.~Morrell, J.S.~Nichols, L.J.~Smith, 
and A.F.J.~Moffat.  My travel to Prague was supported by NASA through 
SAO/Chandra grant GO5-6006D.  The Space Telescope Science Institute is
operated by AURA, Inc., under NASA contract NAS5-26555.


\begin{thebibliography}{}

\bibitem{}
Bohannan, B. and Walborn, N.R.,
\newblock{\em Publications of the A.S.P.}, 101, 520, 1989

\bibitem{}
Conti, P.S. and Leep, E.M.,
\newblock{\em Astrophysical Journal}, 193, 113, 1974

\bibitem{}
Crowther, P.A. and Bohannan, B.,
\newblock{\em Astronomy and Astrophysics}, 317, 532 1997

\bibitem{}
Crowther, P.A. and Smith, L.J.,
\newblock{\em Astronomy and Astrophysics}, 320, 500, 1997

\bibitem{}
Donati, J.-F. et al.,
\newblock{\em Monthly Notices of the R.A.S.}, 333, 55, 2002

\bibitem{}
Donati, J.-F. et al.,
\newblock{\em Monthly Notices of the R.A.S.}, 365, L6, 2006a

\bibitem{}
Donati, J.-F. et al.,
\newblock{\em Monthly Notices of the R.A.S.}, 370, 629, 2006b

\bibitem{}
Gagn\'e, M. et al.,
\newblock{\em Astrophysical Journal}, 628, 986 and 634, 712, 2005

\bibitem{}
Herrero, A. et al.,
\newblock{\em Astronomy and Astrophysics}, 261, 209, 1992

\bibitem{}
Howarth, I.D.,
\newblock{\em IAU Symposium}, 215, 33, 2004

\bibitem{}
Howarth, I.D. and Smith, K.C.,
\newblock{\em Monthly Notices of the R.A.S.}, 327, 353, 2001

\bibitem{}
Hubrig, S. et al.,
\newblock{\em Monthly Notices of the R.A.S.}, 369, L61, 2006

\bibitem{}
Lloyd, C. et al.,
\newblock{\em Publications of the A.S.P.}, 107, 1030, 1995

\bibitem{}
Maeder, A. and Meynet, G.,
\newblock{\em Annual Review of Astronomy and Astrophysics}, 38, 143, 2000

\bibitem{}
Mihalas, D. et al.,
\newblock{\em Astrophysical Journal}, 175, L99, 1972

\bibitem{}
Morrell, N.I. et al.,
\newblock{\em Publications of the A.S.P.}, 117, 699, 2005

\bibitem{}
Naz\'e, Y.,
\newblock{PhD thesis, Universit\'e de Li\`ege}, 2004

\bibitem{}
Naz\'e, Y. et al.,
\newblock{\em Astronomy and Astrophysics}, 372, 195, 2001

\bibitem{}
Negueruela, I. et al.,
\newblock{\em Astronomische Nachrichten}, 325, 749, 2004

\bibitem{}
Nota, A. et al.,
\newblock{\em Astrophysical Journal}, 448, 788, 1995

\bibitem{}
Parker, J.Wm. et al.,
\newblock{\em Astronomical Journal}, 103, 1205, 1992

\bibitem{}
Pasquali, A. et al.,
\newblock{\em Astrophysical Journal}, 478, 340, 1997

\bibitem{}
Pasquali, A. et al.,
\newblock{\em Astronomy and Astrophysics}, 343, 536, 1999

\bibitem{}
Pellerin, A. et al.,
\newblock{\em Astrophysical Journal Supplements}, 143, 159, 2002

\bibitem{}
Rountree, J. and Sonneborn, G.,
\newblock{\em Astrophysical Journal}, 369, 515, 1991

\bibitem{}
Rountree, J. and Sonneborn, G.,
\newblock{\em NASA Reference Publication}, No. 1312, 1993

\bibitem{}
Sana, H. et al.,
\newblock{\em Monthly Notices of the R.A.S.}, 372, 661, 2006

\bibitem{}
Schlesinger, F.,
\newblock{\em Astrophysical Journal}, 33, 260, 1911

\bibitem{}
Smith, M.A. and Fullerton, A.W.,
\newblock{\em Publications of the A.S.P.}, 117, 13, 2005

\bibitem{}
Stahl, O. et al.,
\newblock{\em Astronomy and Astrophysics}, 127, 49, 1983

\bibitem{}
Stahl, O. et al.,
\newblock{\em Astronomy and Astrophysics}, 312, 539, 1996

\bibitem{}
Wade, G.A. et al.,
\newblock{\em Astronomy and Astrophysics}, 451, 195, 2006

\bibitem{}
Walborn, N.R.,
\newblock{\em Astrophysical Journal}, 167, L31, 1971a

\bibitem{}
Walborn, N.R.,
\newblock{\em Astrophysical Journal Supplements}, 23, 257, 1971b

\bibitem{}
Walborn, N.R.,
\newblock{\em Astronomical Journal}, 77, 312, 1972

\bibitem{}
Walborn, N.R.,
\newblock{\em Astronomical Journal}, 78, 1067, 1973

\bibitem{}
Walborn, N.R.,
\newblock{\em Astrophysical Journal}, 205, 419 1976

\bibitem{}
Walborn, N.R.,
\newblock{\em Astrophysical Journal}, 215, 53, 1977

\bibitem{}
Walborn, N.R.,
\newblock{\em Astrophysical Journal Supplements}, 44, 535, 1980

\bibitem{}
Walborn, N.R.,
\newblock{\em Astrophysical Journal}, 256, 452, 1982

\bibitem{}
Walborn, N.R.,
\newblock{\em A.S.P. Conference Series}, 242, 217, 2001

\bibitem{}
Walborn, N.R.,
\newblock{\em A.S.P. Conference Series}, 304, 29, 2003

\bibitem{}
Walborn, N.R.,
\newblock{\em STScI May Symposium}, in press, 2006

\bibitem{}
Walborn, N.R. et al.,
\newblock{\em NASA Reference Publication}, No. 1155, 1985

\bibitem{}
Walborn, N.R. et al.,
\newblock{\em Astrophysical Journal}, 393, L13, 1992

\bibitem{}
Walborn, N.R. et al.,
\newblock{\em NASA Reference Publication}, No. 1363, 1995

\bibitem{}
Walborn, N.R. et al.,
\newblock{\em Publications of the A.S.P.}, 112, 1243, 2000

\bibitem{}
Walborn, N.R. et al.,
\newblock{\em Astronomical Journal}, 123, 2754, 2002a

\bibitem{}
Walborn, N.R. et al.,
\newblock{\em Astrophysical Journal Supplements}, 141, 443, 2002b

\bibitem{}
Walborn, N.R. et al.,
\newblock{\em Astrophysical Journal}, 588, 1025, 2003

\bibitem{}
Walborn, N.R. et al.,
\newblock{\em Astrophysical Journal}, 608, 1028, 2004a

\bibitem{}
Walborn, N.R. et al.,
\newblock{\em Astrophysical Journal}, 617, L61, 2004b

\bibitem{}
Walborn, N.R. and Fitzpatrick, E.L. (WF),
\newblock{\em Publications of the A.S.P.}, 102, 379, 1990

\bibitem{}
Walborn, N.R. and Fitzpatrick, E.L.,
\newblock{\em Publications of the A.S.P.}, 112, 50, 2000

\bibitem{}
Walborn, N.R. and Howarth, I.D.,
\newblock{\em Publications of the A.S.P.}, 112, 1446, 2000

\bibitem{}
Walborn, N.R. and Nichols, J.S.,
\newblock{\em Astrophysical Journal}, 425, L29, 1994

\bibitem{}
Walborn, N.R. and Panek, R.J.,
\newblock{\em Astrophysical Journal}, 280, L27, 1984a

\bibitem{}
Walborn, N.R. and Panek, R.J.,
\newblock{\em Astrophysical Journal}, 286, 718, 1984b

\bibitem{}
Walborn, N.R. and Parker, J.Wm.,
\newblock{\em Astrophysical Journal}, 399, L87, 1992

\end{thebibliography}
\end{document}